\documentclass[12pt]{article}
\usepackage{epsfig, epsf, graphicx, subfigure, color}
\usepackage{pstricks, pst-node, psfrag}
\usepackage{amssymb,amsmath}
\usepackage{verbatim,enumerate}
\usepackage{rotating, lscape}
\usepackage{setspace}
\usepackage{bbm}
\usepackage{natbib}
\usepackage{amsthm}
\usepackage{url}
\usepackage{multirow}
\usepackage{caption}

\theoremstyle{definition}
	
\newtheorem{Definition}{Definition}

\def\cov{\hbox{Cov}}
\newcommand{\bs}{\mathbf{s}}
\newcommand{\bh}{\mathbf{h}}
\newcommand{\bl}{\mathbf{0}}

\begin{document}
	\thispagestyle{empty} \baselineskip=28pt \vskip 5mm
	\begin{center} {\Large{\bf Visualization and Assessment of Spatio-temporal Covariance Properties}}
	\end{center}
	
	\baselineskip=12pt \vskip 5mm
	
	\begin{center}\large
		Huang Huang\footnote[1]{
			\baselineskip=10pt CEMSE Division, King Abdullah University of Science and Technology, Thuwal 23955-6900 Saudi Arabia.
			E-mail: huang.huang@kaust.edu.sa; ying.sun@kaust.edu.sa
		} and Ying Sun$^1$\end{center}
	
	\baselineskip=16pt \vskip 1mm \centerline{\today} \vskip 8mm
	
	\begin{center}
		{\large{\bf Abstract}}
	\end{center}
	\baselineskip=17pt
			
	Spatio-temporal covariances are important for describing the spatio-temporal variability of underlying random processes in geostatistical data.
	For second-order stationary processes, there exist subclasses of covariance functions that assume a simpler spatio-temporal dependence structure with separability and full symmetry.
	However, it is challenging to visualize and assess separability and full symmetry from spatio-temporal observations. 
	In this work, we propose a functional data analysis approach that constructs test functions using the cross-covariances from time series observed at each pair of spatial locations.
	These test functions of temporal lags summarize the properties of separability or symmetry for the given spatial pairs. We use functional boxplots to visualize the functional median and the variability of the test functions, where the extent of departure from zero at all temporal lags indicates the degree of non-separability or asymmetry. We also develop a rank-based nonparametric testing procedure for assessing the significance of the non-separability or asymmetry.
	The performances of the proposed methods are examined by simulations with various commonly used spatio-temporal covariance models.
	To illustrate our methods in practical applications, we apply it to real datasets, including weather station data and climate model outputs.
	\begin{doublespace}
		
		\par\vfill\noindent
		{\bf Some key words}:  full symmetry, functional boxplot, functional data ranking, rank-based test, separability, spatio-temporal covariance
		\par\medskip\noindent
	\end{doublespace}
	
	\clearpage\pagebreak\newpage \pagenumbering{arabic}
	\baselineskip=26.5pt
	

\section{Introduction}
Spatio-temporal covariance functions play an important role in parametrically modeling geostatistical data, particularly for Gaussian random fields. Spatio-temporal models often have complex structures and rely on various assumptions to simplify the model and reduce the computational burden~\citep{Gneiting2007}, such as full symmetry and separability~\citep{Cressie1999,Kyriakidis1999,Gneiting2002,Stein2005}.
Specifically, let $Z(\mathbf{s},t)$ be a second-order stationary random process at spatial location $\mathbf s\in\mathbb{R}^d$, and temporal point $t\in\mathbb{R}.$ The covariance function $C(\bh,u)=\cov\{Z(\mathbf{s},t),Z(\mathbf s+\bh,t+u)\}$ is positive definite and only depends on the spatial lag $\mathbf{h}$ and temporal lag $u$.
Then, the covariance function is called fully symmetric if
$C(\bh,u)=C(-\bh,u)=C(\bh,-u)$, and separable if
$C(\bh,u)=C(\bh,0)C(\bl,u)/C(\bl,0)$, for any spatial lag $\mathbf{h}$ and temporal lag $u$. One can show that separability implies full symmetry.
The separable covariance model is a product of purely spatial and temporal covariances, ignoring the interaction between space and time;
the symmetric covariance model acknowledges the interaction but assumes that the spatio-temporal cross-covariances depend on neither the direction of temporal lags nor the direction of spatial lags.
Although these models with simplified structures are easier to fit, they have limitations and may not be realistic in practical applications.
In separable models, even small spatial changes can lead to large changes in the correlations, causing a lack of smoothness~\citep{Stein2005}, and full symmetry is often violated when data are influenced by dynamic processes with a prevailing flow direction~\citep{Gneiting2002}.

Therefore, it has been an active research area to develop more flexible spatio-temporal covariance models motivated by real applications. \citet{Jones1997} developed families of spectral densities that lead to non-separable covariance models without a closed form. \citet{Cressie1999} proposed families of non-separable covariance functions based on the Fourier transform of non-negative, finite measures with explicit expressions, but these are limited to classes with known analytical solutions of Fourier integrals. Later, \citet{Gneiting2002} extended this approach to more general classes with Fourier-free implementations. 
All these models are non-separable but fully symmetric. \citet{Stein2005} generated covariance functions that are isotropic but not fully symmetric by taking derivatives of fully symmetric models. \citet{Gneiting2007} proposed anisotropic, asymmetric models by adding a compactly supported Lagrangian correlation function to a fully symmetric covariance model, where the coefficient of the Lagrangian component controls the extent of asymmetry.

In real data analysis, before choosing a covariance function from the existing models, it is necessary to assess the properties of separability and symmetry either by exploratory data analysis or through formal hypothesis tests. However, it is challenging to visualize and assess such properties from spatio-temporal observations. 
\citet{Brown2001} determined the separability from the closeness to zero of the maximum likelihood estimates of blurring parameters associated with separability, without considering the uncertainty. \citet{Shitan1995} developed an asymptotic test for separability, but it is restricted to spatial autoregressive processes. \citet{Fuentes2006} presented a test of separability in the spectral domain using a simple two-factor analysis of variance. 
Regarding likelihood-based tests, \citet{Mitchell2006} proposed a likelihood ratio test of separability for replicated multivariate data that examines whether the covariance matrix is a Kronecker product of two matrices of smaller dimensions. \citet{Mitchell2005} extended this approach to test the separability of spatio-temporal covariances by partitioning the observations into approximate replicates.  After \citet{Scaccia2005} presented tests of symmetry and separability for lattice processes using the periodogram, \citet{Li2007} developed a unified testing framework for both separability and symmetry by constructing contrasts of covariances at selected spatio-temporal lags.

Existing methods mostly build separability and symmetry tests based on given spatial and temporal lag sets, and work on visualization is sparse. 
In this paper, we propose formulating separability and symmetry test functions as functional data, and provide a visualization tool to illustrate these properties. 
These test functions are functions of temporal lags and constructed from the cross-covariances of each pair of the time series observed at spatial locations.
With the obtained functional data of test functions, we develop rank-based testing procedures that are model-free with data depth-based functional data ranking techniques. Because the test statistics are rank-based, the results are more robust to outliers, which may come from erroneous measurements of the variables under study or poor estimates of the covariances due to limited sample size.

The rest of the paper is organized as follows. In \S\ref{sec:method}, we introduce separability and symmetry test functions, which are used to visualize the covariance property of separability and symmetry, and provide the detailed rank-based testing procedure. In \S\ref{sec:simulation}, we illustrate the visualization tools for simulated data and demonstrate the rank-based test for various spatio-temporal covariance models. In \S\ref{sec:application}, we apply our methods to two real datasets, wind speed from monitoring stations, and surface temperatures and wind speed from numerical model outputs, to study their covariance structures.

\section{Methodology}
\label{sec:method}
\subsection{Non-separable covariance functions}
\label{subsec:sep1}
\citet{Gneiting2002} introduced the following rich family of non-separable spatio-temporal covariance functions, which includes separable ones as a special case,
$$
\label{eq:sepPara}
C(\bh,u)=\frac{\sigma^2}{(a|u|^{2\alpha}+1)^\beta}
\hbox{exp}\Big\{-\frac{c\|\bh\|^{2\gamma}}{(a|u|^{2\alpha}+1)^{\beta\gamma}}\Big\},
$$
where $\sigma^2$ is the variance, $a, c\geq0$ determine the temporal and spatial range, $\alpha, \gamma\in[0,1]$ control the temporal and spatial smoothness, and $\beta\in[0,1]$ is the spatio-temporal interaction.
The covariance function is separable when $\beta=0$, and a larger $\beta$ is associated with a more non-separable model. 

\citet{Cressie1999} also proposed a non-separable model,
$$
C(\bh,u)=\frac{\sigma^2(a|u|+1)}{\{(a|u|+1)^2+b^2\|\bh\|^2\}^{3/2}},
$$
and a possible corresponding separable model was discussed in~\citet{Mitchell2005} as,
$$
C(\bh,u)=\frac{\sigma^2}{(a|u|+1)^2(b^2\|\bh\|^2+1)^{3/2}}.
$$

Later, \citet{Rodrigues2010} further defined the properties of positive and negative non-separability and showed that the two aforementioned models are both positively non-separable. To investigate models of negative non-separability, we consider the one from \citet{Cesare2001},
$$
C(\bh,u)=k_1C_s(\bh)C_t(u)+k_2C_s(\bh)+k_3C_t(u),
$$
where $k_1>0$, $k_2,k_3\geq0$, and $C_s(\cdot)$ and $C_t(\cdot)$ are valid spatial and temporal covariance functions, respectively.

Table~\ref{tab:sepModels} summarizes the specific models used to illustrate our methods, after we plugged in the appropriate parameter values, chose suitable building covariance functions, and applied some trivial transformations, as necessary.

\begin{table}[h!]
	\centering
	\caption{\label{tab:sepModels}Notations of the chosen covariance models.}
	\begin{tabular}{|c|c|c|}
		\hline
		\bf Model& \bf Notation& \bf Formula\\\hline
		Gneiting&$C^G(\bh,u;\beta)$& $\dfrac{1}{0.5|u|+1}
		\hbox{exp}\Big\{-\dfrac{\|\bh\|}{(0.5|u|+1)^{\beta/2}}\Big\}$\\\hline
		\begin{tabular}{@{}c@{}}Cressie and Huang \\ separable\end{tabular} &$C^{CH}_{sep}(\bh,u)$&$\dfrac{1}{(0.5|u|+1)^2(\|\bh\|^2+1)^{3/2}}$\\\hline
		\begin{tabular}{@{}c@{}}Cressie and Huang \\ non-separable\end{tabular} &$C^{CH}_{nsep}(\bh,u)$&$\dfrac{(0.5|u|+1)}{\{(0.5|u|+1)^2+\|\bh\|^2\}^{3/2}}$\\\hline	
		Cesare&$C^C(\bh,u;k)$&$\dfrac{1}{3}\big\{\exp(-\bh)\exp(-u)+\exp(-u)+k\exp(-\bh)\big\}$\\\hline
	\end{tabular}
\end{table}

\subsection{Visualization for separability}
\label{subsec:sep2}
Noting that separable covariance functions require
$C(\bh,u)=C(\bh,0)C(\bl,u)/C(\bl,0)$, for any spatial lag $\mathbf{h}$ and temporal lag $u$, or equivalently, $C(\bh,u)/C(\bh,0)$ remains a constant $C(\bl,u)/C(\bl,0)$ for any given $\bh$. Therefore, we introduce the separability test function defined below.
\begin{Definition}
	\label{def:sep}
	Given a valid spatio-temporal covariance function $C(\bh,u)$, the separability test function $f_\bh(u)$ is a function of temporal lag $u$ for any spatial lag $\bh$, defined as
	\[f_\bh(u)=C(\bh,u)/C(\bh,0)-C(\bl,u)/C(\bl,0).\]
\end{Definition}
By Definition~\ref{def:sep}, $f_\bh(u)$ is $0$ for any $u$ and $\bh$ if $C(\bh,u)$ is separable, and moves away from zero for non-separable models. Suppose $f_{\bh_1}(u),\ldots,f_{\bh_n}(u)$ are a set of test functions for pairs of locations with spatial lags $\bh_1,\ldots,\bh_n$. We use functional boxplots~\citep{Sun2011} to visualize these separability test functions. The motivation is that functional boxplots can characterize the most representative functional realization, or the functional median, and remove the outliers, which are often involved in the datasets, especially when we are going to use sample covariances to estimate the test functions. 
Functional boxplots order the functional observations by their functional data depths, identify the observation with the largest depth value as the functional median and construct the $50\%$ central region that covers half of the data with largest depth values. Although depth values in the functional boxplot are calculated by the modified band depth proposed by~\citep{Lopez-Pintado2009} by default, it is also possible to choose other depth notions developed for functional data, such as integrated data depth~\citep{Fraiman2001}, half-region depth~\citep{Lopez-Pintado2011}, and extremal depth~\citep{Narisetty2015}. Outliers are detected by the $1.5$ times the $50\%$ central region empirical rule. The factor $1.5$ can be adjusted for outlier detection proposes in an adjusted functional boxplot, but the adjustment is not necessary for visualizing functional data (\cite{Sun2012a}). Therefore, it is worth pointing out that we use the functional boxplot for visualizing covariance properties in this paper rather than detecting outliers. Alternatively, contour plots are the traditional way to visualize the pattern of covariances with respect to different spatial and temporal lags. 
We show next that the functional boxplots provide a much better visualization and interpretation of the separability than the contour plots.

\begin{figure}[ht!]
	\centering
	\includegraphics[width=\textwidth]{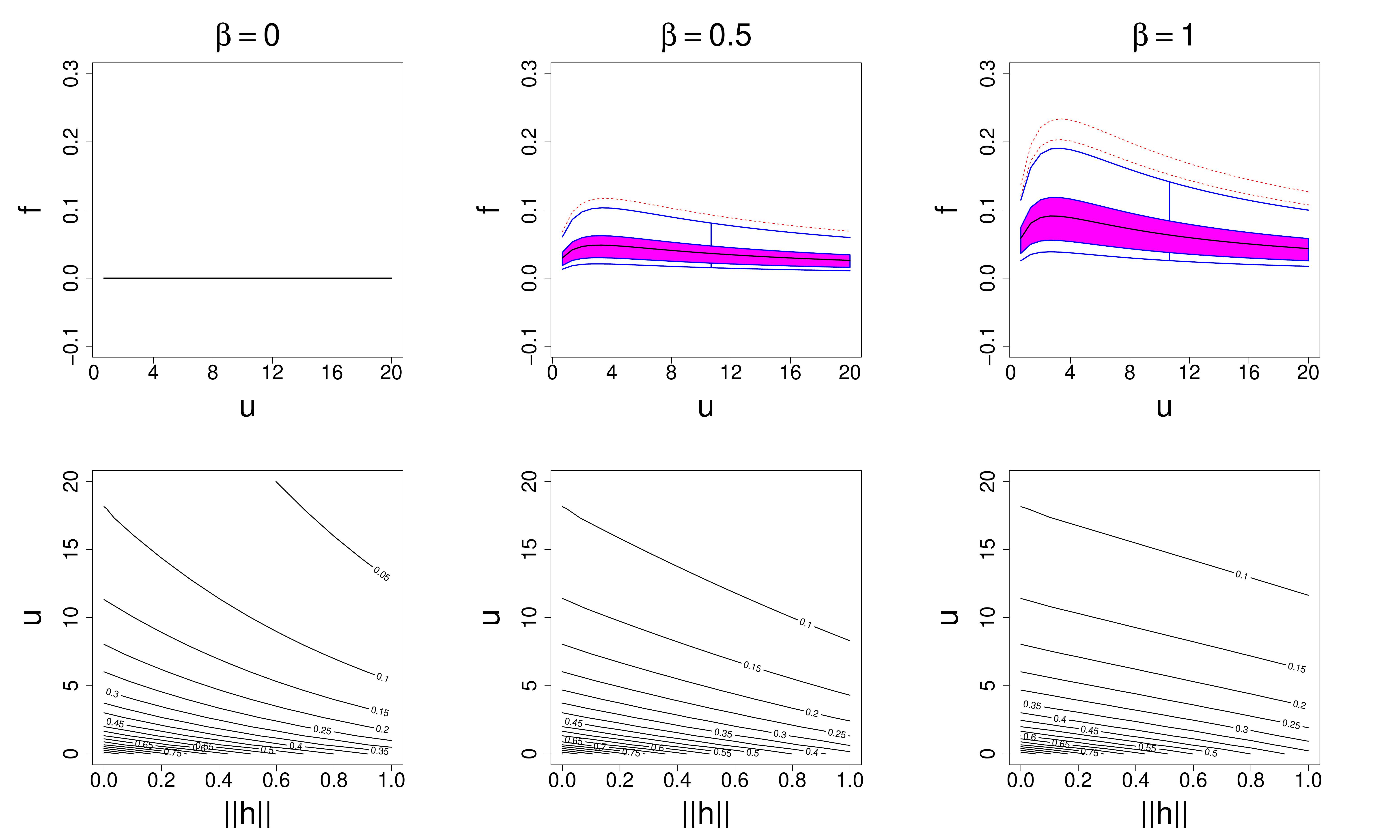}
	\caption{\label{fig:sep1}
		Top: functional boxplots of the separability test functions for the Gneiting model $C^G(\bh,u;\beta)$ with different values of $\beta$. The magenta area is the central region, the black line is the median, and the red lines are outliers.
		Bottom: contour plots of the covariances for the Gneiting model $C^G(\bh,u;\beta)$ with different values of $\beta$. }
\end{figure}

Firstly, we use functional boxplots of the separability test functions and contour plots of the covariances to visualize the Gneiting model $C^G(\bh,u;\beta)$ for different values of $\beta$. 
The spatial locations are on a $4\times4$ regular grid within the unit square $[0,1]\times[0,1]$. The top panels in Figure~\ref{fig:sep1} show the functional boxplots applied to the separability test functions for the Gneiting model $C^G(\bh,u;\beta)$ with different values of $\beta$. Each curve is associated with a specific pair of spatial locations. The central region (magenta area) and the median (black curve) move away from zero as $\beta$ increases. 
The bottom panels in Figure~\ref{fig:sep1} show the contours of the covariances for different spatial and temporal lags. 
In the separable case, the covariances at non-zero temporal lags should decay more rapidly as the spatial lag increases. 
We do see this trend in the contour plots; however, the separability is much more obvious in the functional boxplot.

\begin{figure}[ht!]
	\centering
	\includegraphics[width=\textwidth]{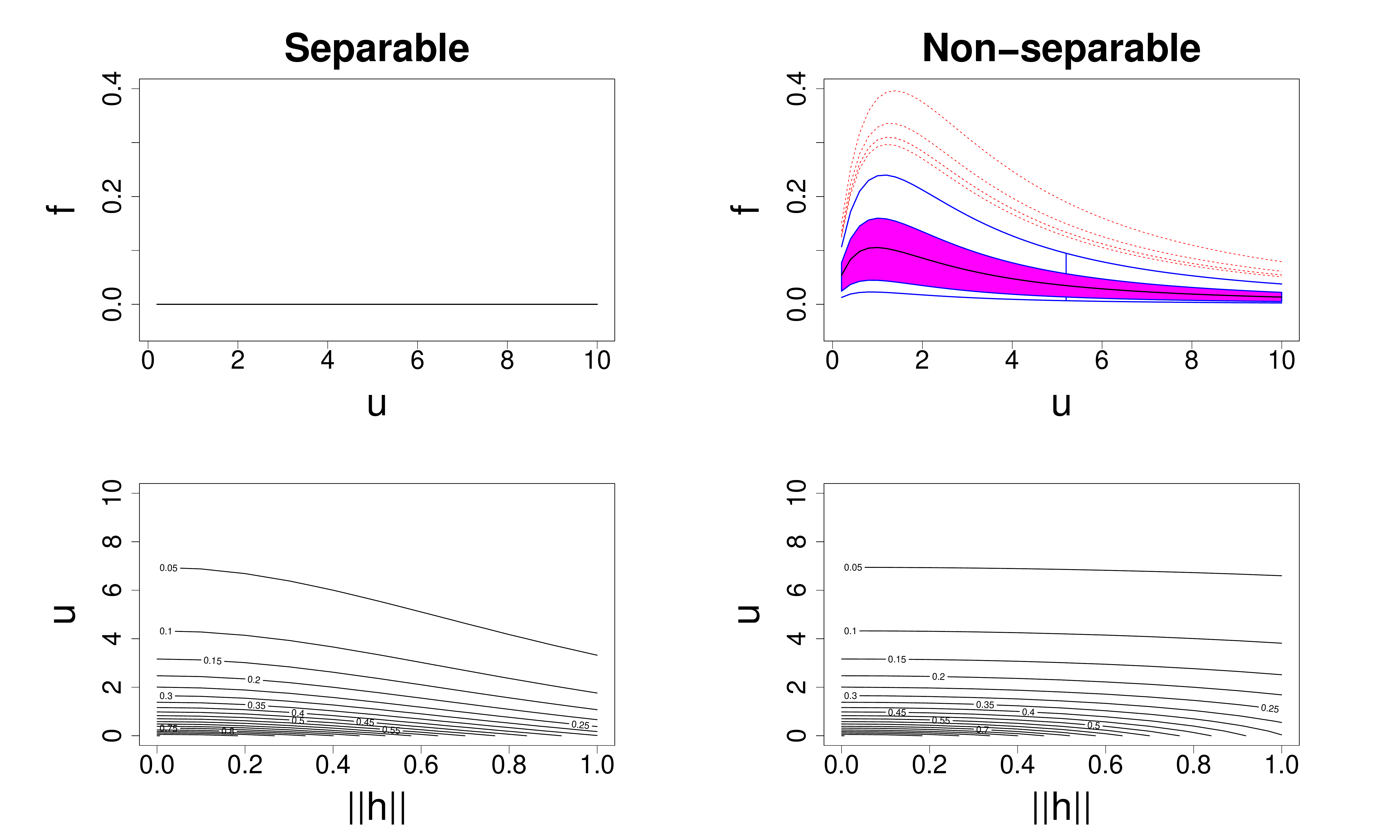}
	\caption{\label{fig:sep2}
		Top: functional boxplots of the separability test functions for the Cressie and Huang separable model $C^{CH}_{sep}(\bh,u)$ and non-separable model $C^{CH}_{nsep}(\bh,u)$. The magenta area is the central region, the black line is the median, and the red lines are outliers.
		Bottom: contour plots of the covariances for $C^{CH}_{sep}(\bh,u)$ and $C^{CH}_{nsep}(\bh,u)$. }
\end{figure}

\begin{figure}[ht!]
	\centering
	\includegraphics[width=\textwidth]{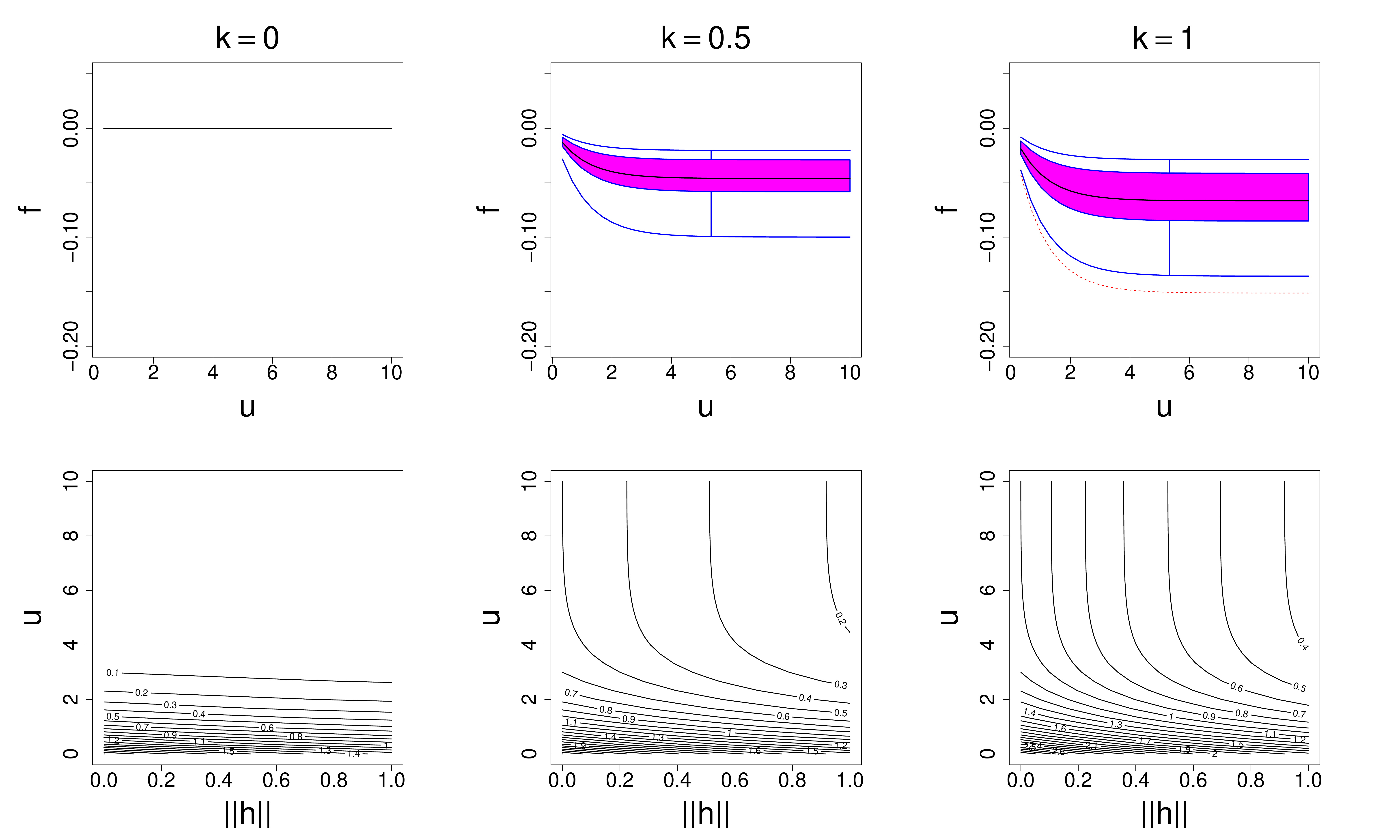}
	\caption{\label{fig:sep3}
		Top: functional boxplots of the separability test functions for the Cesare model $C^C(\bh,u;k)$ with different values of $k$. The magenta area is the central region, the black line is the median, and the red lines are outliers.
		Bottom: contour plots of covariances for $C^C(\bh,u;k)$ with different values of $k$. }
\end{figure}

Similarly, we draw the same plots to compare the Cressie and Huang separable model $C^{CH}_{sep}(\bh,u)$ and the Cressie and Huang non-separable model $C^{CH}_{nsep}(\bh,u)$, and the Cesare model $C^C(\bh,u;k)$ for different values of $k$. The results are shown in Figures~\ref{fig:sep2} and \ref{fig:sep3}.
The functional boxplots clearly show the separability or non-separability of each test function, and, if non-separable, the positive or negative non-separability.
While the contour plots show different patterns and the separable covariance decays much faster than non-separable covariances at non-zero temporal lags along an increasing spatial lag, the indication for separability is still unclear.

\subsection{Visualization and rank-based testing procedure for real data}
\label{subsec:real}
In \S\ref{subsec:sep1}--\ref{subsec:sep2}, we discussed the visualization of true spatio-temporal covariance functions, and we saw that the functional boxplots produce intuitive indications of separability in the test functions.
However, the true covariance model is unknown in practice, and we must use sample covariance estimators to estimate the test functions without assuming any parametric model to fit.
For each pair of spatial locations, we use the two time series to estimate all the components in the separability test function $f_\bh(u)$ by sample covariances.
In other words, if we have observations $Z(\bs,t)$ at $p$ time points $1,\ldots,p$, at two locations $\mathbf{s}_1$ and $\mathbf{s}_2$ with a spatial lag $\bh=\mathbf{s}_2-\mathbf{s}_1$, then the test function is estimated by
\[
\widehat{f}_{\bh}(u)=\widehat{C}(\bh,u)/\widehat{C}(\bh,0)-\widehat{C}(\bl,u)/\widehat{C}(\bl,0),\quad u=0,\ldots,p-1, 
\]
where
{\small \[\widehat{C}(\bh,u)=\dfrac{\displaystyle\sum_{i=1}^{p-u}\big\{Z(\mathbf{s}_1,i)-\frac{1}{p-u}\sum_{j=1}^{p-u}Z(\mathbf{s}_1,j)\big\}\big\{Z(\mathbf{s}_1+\bh,i+u)-\frac{1}{p-u}\sum_{j=1}^{p-u}Z(\mathbf{s}_1+\bh,j+u)\big\}}{p-u},\]}
and other quantities are calculated similarly.

With these estimated separability test functions, we also develop a rank-based test for the separability hypothesis. \citet{Lopez-Pintado2009} proposed a test for functional data to decide whether two sets of samples are from the same distribution. It can be viewed as a functional data generalization of the univariate Wilcoxon rank test and has many applications. For example, \cite{Sun2012b} proposed a robust functional analysis of variance (ANOVA) model for testing whether the differences in functional data observed from multiple groups are significant. The main idea is to see whether the positions of the samples from the two sets are similar, with respect to a reference distribution.
We follow a similar procedure to test the null hypothesis, separability, by comparing the given observations to a simulated dataset under $H_0$: separability. In this rank-based test, two additional datasets are required. One follows the distribution under the null hypothesis $H_0$, and the other is a reference dataset following the distribution under the null hypothesis $H_0$ or the same distribution as the observations. 
We propose to generate these two datasets under $H_0$ by simulating data from a constructed separable covariance function $\widehat{C}^{H_0}$, for which we estimate the covariance at observed spatial and temporal lags by sample covariances, respectively, and let $\widehat{C}^{H_0}(\bh,u)=\widehat{C}(\bh,0)\widehat{C}(0,u)$.
After we get all estimated separability test functions, the hypothesis test can be performed to compare the two populations. More details can be found in the paper by \citet{Lopez-Pintado2009}, and we briefly explain the procedure as follows:
 
\begin{enumerate}
	\item[Step 1] Estimate the test functions for the observations, denoted by $\widehat{f}_{\bh_i}(u)$, $i=1,\ldots,n$.
	\item[Step 2] Generate two sets of spatio-temporal datasets from the constructed separable covariance function $C^{H_0}(\bh,u)$ and estimate the test functions, denoted by $\widehat{f}^0_{\bh_j}(u)$, $j=1,\ldots,m$ and $\widehat{f}^R_{\bh_l}(u)$, $l=1,\ldots,r$.
	\item[Step 3] For each $i$, combine $\widehat{f}_{\bh_i}(u)$ and all $\widehat{f}^R_{\bh_l}(u)$ for $l=1,\ldots,r$, and compute their functional band depths and modified band depths~\citep{Lopez-Pintado2009}. Order these functions by their band depth values, and use modified band depths to separate ties. 
	Let $r_i$ be the proportion of functions $\widehat{f}^R_{\bh_l}(u)$ with smaller depth values than $\widehat{f}_{\bh_i}(u)$. 
	\item[Step 4] Repeat Step 3 for each $\widehat{f}^0_{\bh_j}(u)$, and let $r'_j$ be the proportion of functions $\widehat{f}^R_{\bh_l}(u)$ with smaller depth values than  $\widehat{f}_{\bh_i}(u)$. 
	\item[Step 5] Order $r_1,\ldots,r_n,r'_1,\ldots,r'_m$ from smallest to largest, and assume the ranks of $r_1,\ldots,r_n$ are $q_1,\ldots,q_n$. Then the test statistics is $W=\sum_{i=1}^{n}q_i$.
\end{enumerate}

Critical values can be determined from the asymptotic distribution of the test statistics, which is the sum of $n$ numbers that are randomly chosen from $1,\ldots,n+m$. In practice, however, we found that the distribution with a limited sample size is not close enough to the asymptotic one. Hence, we use the bootstrap to calculate the critical values.
Under $H_0$, we generate $b$ sets of spatio-temporal datasets from $\widehat{C}^{H_0}(\bh,u)$.  We repeat Steps 1--5 and obtain the test statistics for each dataset, which provides an approximate distribution under the null hypothesis. We examine the performance of this technique through simulations in \S\ref{sec:simulation}, where we use $b=100$.

\subsection{Symmetric covariance functions and visualization}
\citet{Gneiting2007} proposed the following family of asymmetric covariance models that include symmetric ones as a special case,
$$
C(\bh;u)=\frac{(1-\lambda)}{a|u|^{2\alpha}+1}
\hbox{exp}\left\{-\frac{c\|\bh\|}{(a|u|^{2\alpha}+1)^{\beta/2}}\right\}+
\lambda\left(1-\frac{1}{2v}|h_1- v u|\right)_+,$$
where $(\cdot)_+=\max(0,\cdot)$, $h_1$ is the first component of the spatial lag $\mathbf{h}$, $\lambda$ determines the symmetry, $v$ is the velocity along the direction of $h_1$, $a$ and $c$ correspond to the temporal and spatial ranges, $\alpha$ corresponds to smoothness, and $\beta$ corresponds to the separability.
After plugging in appropriate parameter values, we get the following covariance model to study symmetry, referred as the Gneiting model $C^{G'}(\bh,u;\lambda)$.
\[
C^{G'}(\bh,u;\lambda)=\frac{1-\lambda}{0.2|u|+1}
\hbox{exp}(-\|\bh\|)+
\lambda\left(1-\frac{1}{2}|h_1-0.2u|\right)_+.
\]

\begin{figure}[ht!]
	\centering
	\includegraphics[width=\textwidth]{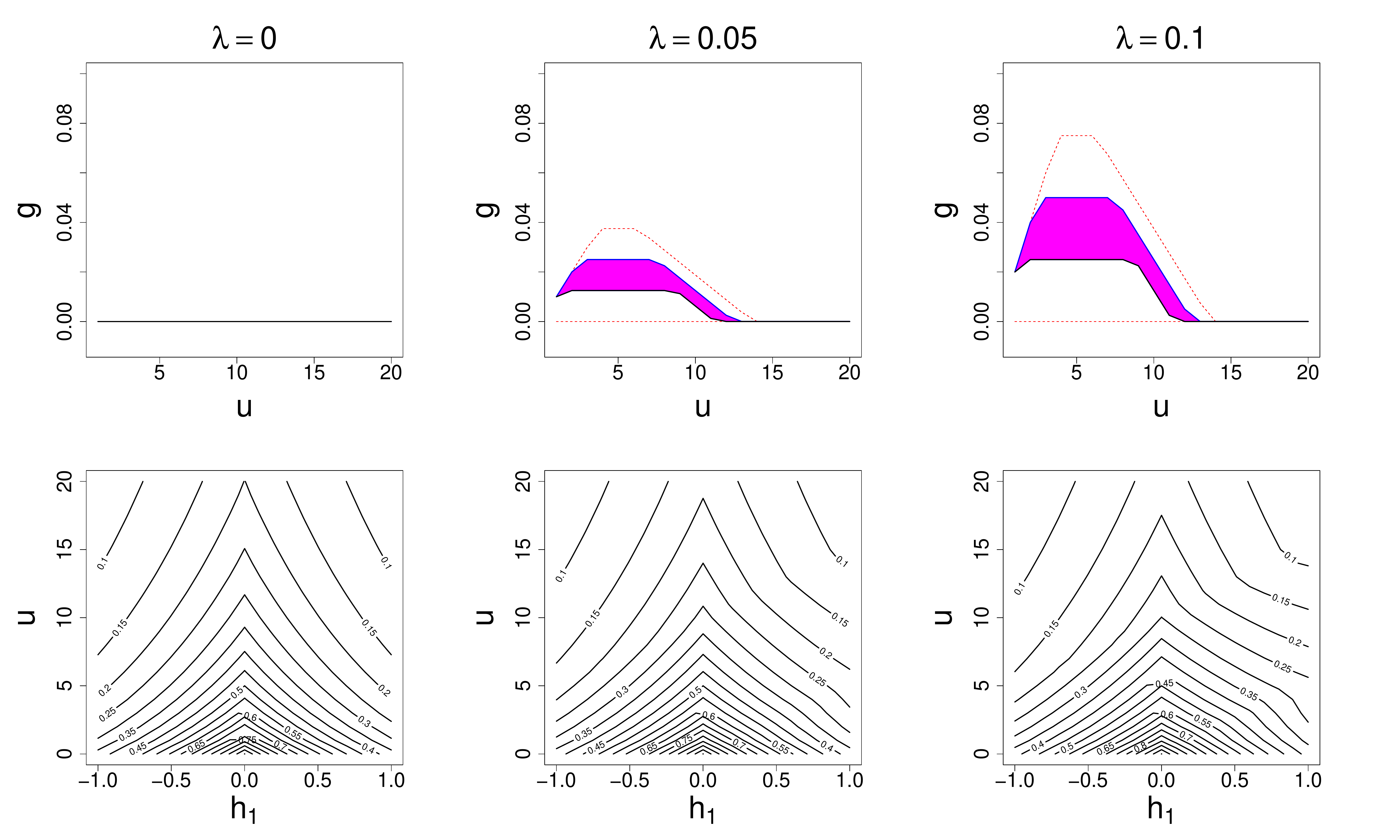}
	\caption{\label{fig:symm}Top: functional boxplots of the symmetry test function for the Gneiting model $C^{G'}(\bh,u;\lambda)$ with different values of $\lambda$. The magenta area is the central region, the black line is the median, and the red lines are outliers. Bottom: contour plots of covariances for $C^{G'}(\bh,u;\lambda)$ with different values of $k$ where $h_2$ is fixed to 0. }
\end{figure}

We know that a fully symmetric covariance function satisfies $C(\bh,u)=C(\bh,-u)$. Analogously to Definition~\ref{def:sep}, we introduce the symmetry test functions with definition below.
\begin{Definition}
	Given a valid spatio-temporal covariance model $C(\bh,u)$, the symmetry test function $g_\bh(u)$ is a function of temporal lag $u$ for any spatial lag $\bh$, defined as
	\[g_\bh(u)=C(\bh,u)-C(\bh,-u).\] 
\end{Definition}
It is easy to see that $g_\bh(u)$ remains zero when the underlying covariance model is fully symmetric. An illustration is given in Figure~\ref{fig:symm} for the Gneiting model $C^{G'}$ with different values of $\lambda$, where we see the symmetry test function $g_\bh(u)$ moves away from zero as $\lambda$ increases. Contour plots of the covariances are also shown in the bottom panels of Figure~\ref{fig:symm}. Since the covariance relies on the temporal lag $u$ and both the first and second components $h_1, h_2$ of the spatial lag $\mathbf{h}$, we fix $h_2=0$, and draw the contours for different values of $u$ and $h_1$.
It is easy to observe that the covariances are clearly asymmetric when $\lambda>0$. For real data, the estimated symmetry test functions are used in the functional boxplots, and a similar rank-based testing procedure described in \S\ref{subsec:real} is developed to test the significance of the asymmetry.

\section{Simulation}
\label{sec:simulation}
\subsection{Simulation design}
Spatio-temporal data collected from monitoring sites are often sparse in space but dense in time. In this simulation, we consider data from $4\times4=16$ regularly spaced locations in the unit square $[0,1]\times[0,1]$, and $p=2,000$ (a comparable number to many real applications) equally spaced time points.

Noting that the correlation between observations with temporal lags greater than $100$ is negligible in each model, we generate our spatio-temporal data sequentially in time. Let $\mathbf{Z}_i$ be a matrix of size $16\times100$, indicating the spatio-temporal observations from time point $(i-1)\times100+1$ to $i\times100$ at the $16$ locations, for $i=1,\ldots,20$. We first generate $\mathbf{Z}_1$, and then generate $\mathbf{Z}_i$, $i=2,\ldots,20$, only conditioning on $\mathbf{Z}_{i-1}$. Thus the entire generated spatio-temporal data is  $\mathbf{Z}=(\mathbf{Z}_1,\mathbf{Z}_2,\ldots,\mathbf{Z}_{20})$.

\subsection{Visualization and assessment for separability}
\begin{figure}[ht!]
	\centering
	\includegraphics[width=\textwidth]{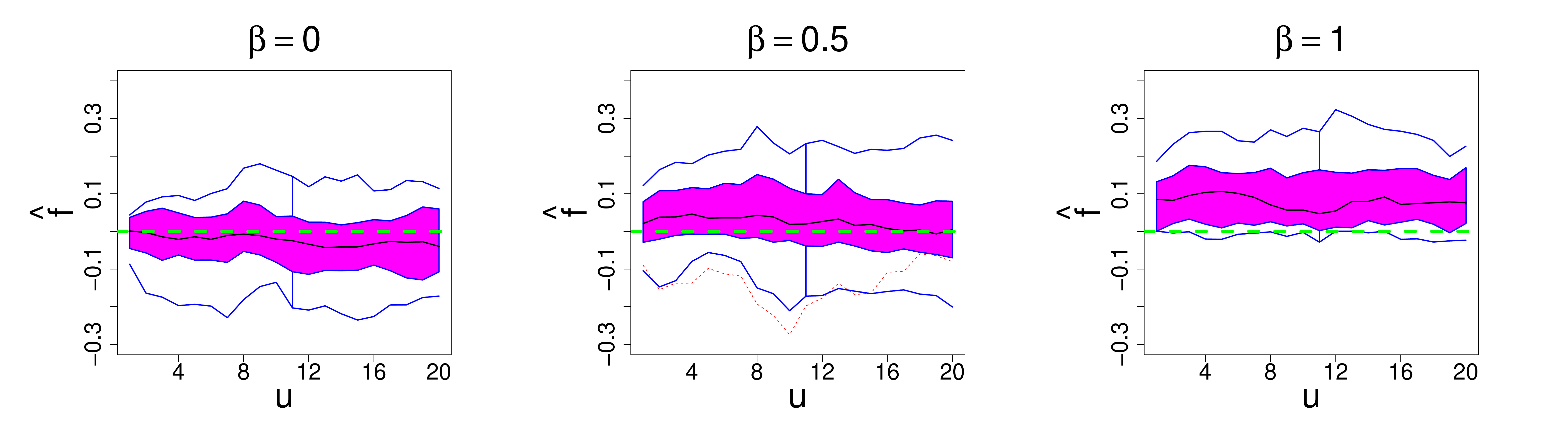}
	\caption{\label{fig:simSep}Functional boxplots of the estimated separability test function for data generated from the Gneiting model $C^G(\bh,u;\beta)$ with $\beta=0$, 0.5, and 1. The magenta area is the central region, the black line is the median, the red lines are outliers, and the green dash line is the reference for separability in theory.}
\end{figure}

The Gneiting model $C^G(\bh,u;\beta)$ with five different values of $\beta=0, 0.25$, $0.5$, $0.75$, and $1$ is used to generate observations $\mathbf{Z}$. 
Three examples of the estimated separability test functions are shown in Figure~\ref{fig:simSep}. 
We see that, in the functional boxplot, the distance between the median (black line) and zero (green dashes) indicates the degree of separability. As $\beta$ increases, the majority of the separability test functions move away from zero.
This gives us a clear visualization of the separability.

\begin{table}[h!]
	\centering
	\caption{\label{tab:power}Type I error and power for simulated data from the Gneiting model $C^G(\bh,u;\beta)$ with different values of $\beta$ in $100$ experiments.}
	\begin{tabular}{c|c|c|c|c|c|c|}
		&\begin{tabular}{c}Nominal\\ Level\end{tabular} &Type I error&\multicolumn{4}{|c|}{Power}\\\cline{2-7}
		&$\alpha$&$\beta=0$&$\beta=0.25$&$\beta=0.5$&$\beta=0.75$&$\beta=1$\\\hline
		Proportion of &$5\%$&0.04&0.26&0.63&0.96&1.00\\\cline{2-7}
		rejections&$1\%$&0.02&0.10&0.45&0.85&0.98\\\hline	
	\end{tabular}
\end{table}

To formally test the separability and examine the performance of the proposed rank-based testing procedure in \S\ref{subsec:real}, we show the Type I error and power for simulated data over $100$ simulations. The results are shown in Table~\ref{tab:power}. We see the Type I error is close to the nominal level of the hypothesis test, and the power for non-separable models increases with an increasing $\beta$. The power is quite close to one when $\beta$ approaches $1$.

\begin{figure}[ht!]
	\centering
	\includegraphics[width=\textwidth]{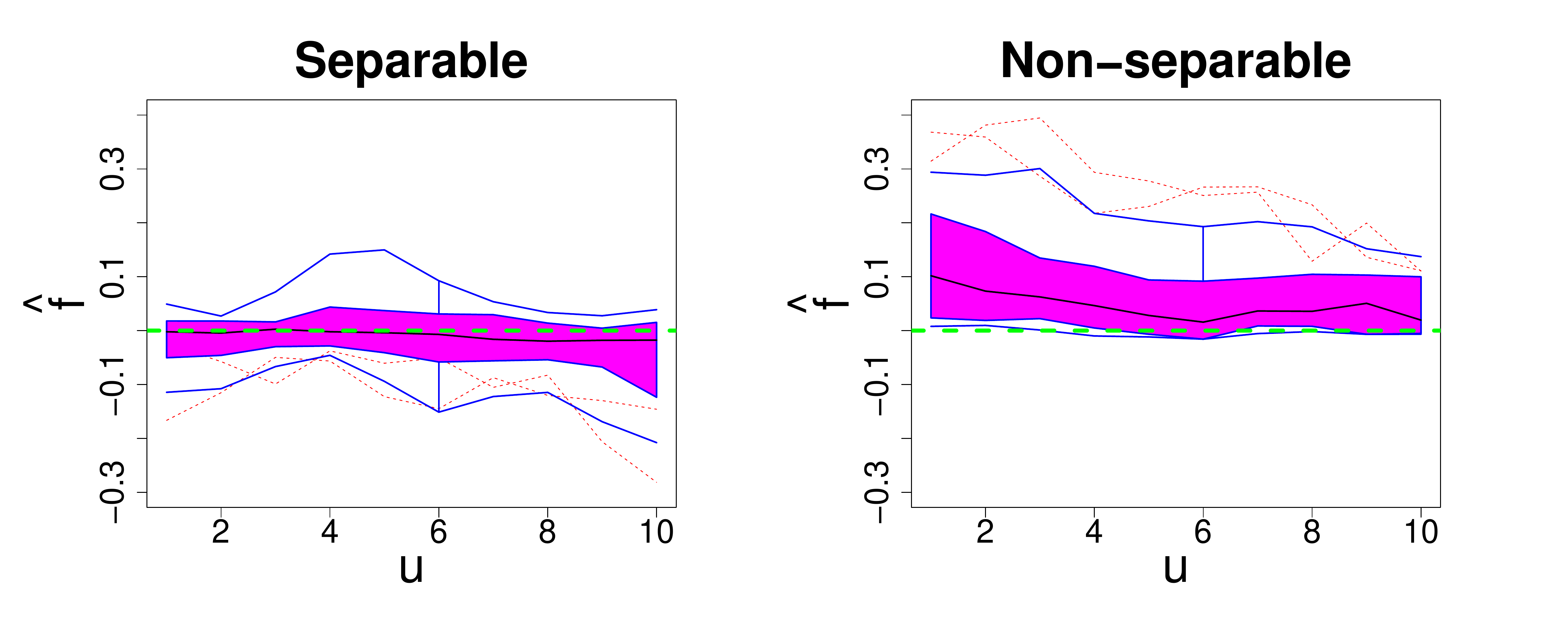}
	\caption{\label{fig:simCH}Functional boxplot of the estimated separability test function for data generated from Cressie-Huang models $C^{CH}_{nsep}$ and $C^{CH}_{sep}$. The magenta area is the central region, the black line is the median, the red lines are outliers, and the green dash line is the reference for separability in theory.}
\end{figure}

In addition, we also take a look at the Cressie-Huang models $C^{CH}_{nsep}$ and $C^{CH}_{sep}$. Estimated separability test functions from one simulated dataset are shown in Figure~\ref{fig:simCH}. We see the majority of the separability test functions from the non-separable model move away from zero. Similarly, a rank-based test is used to examine the performance; the results are shown in Table~\ref{tab:CH}.

\begin{table}[h!]
	\centering
	\caption{\label{tab:CH}Proportion of rejections for simulated data from the Cressie-Huang separable and non-separable models in $100$ experiments.}
	\begin{tabular}{c|c|c|c|}
		&Nominal Level&\multicolumn{2}{|c|}{Model}\\\cline{2-4}
		&$\alpha$&$C^{CH}_{sep}$&$C^{CH}_{nsep}$\\\hline
		Proportion of&$5\%$&0.04&0.97\\\cline{2-4}
		rejections&$1\%$&0.00&0.87\\\hline		
	\end{tabular}
\end{table}

\subsection{Visualization and assessment for symmetry}
\begin{figure}[ht!]
	\centering
	\includegraphics[width=\textwidth]{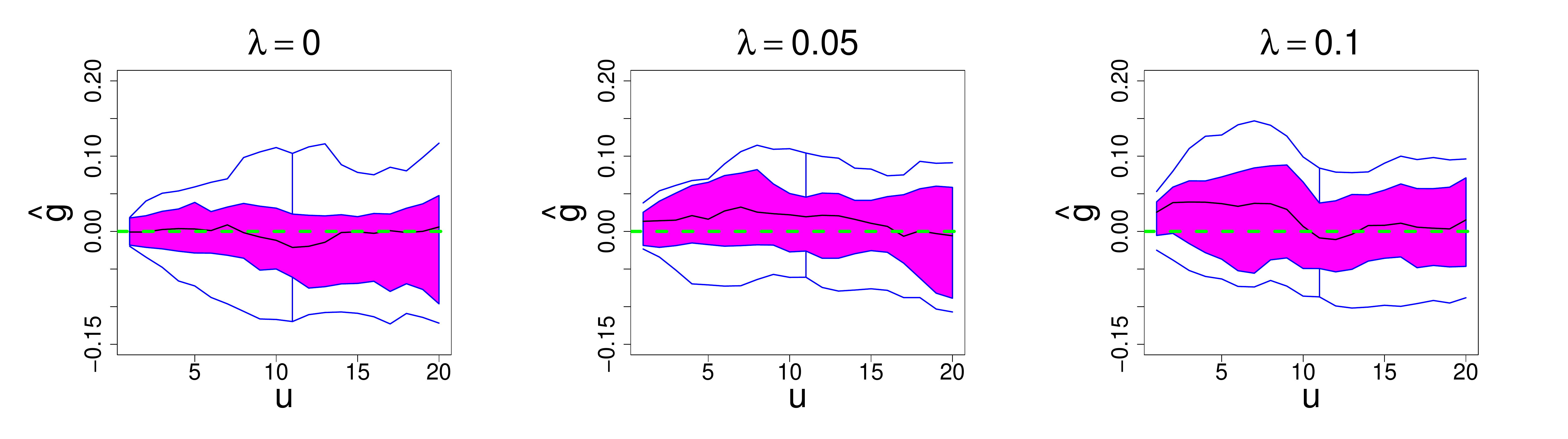}
	\caption{\label{fig:symmSim}Functional boxplots of the estimated symmetry test function for data generated from the Gneiting model $C^{G'}(\bh,u;\lambda)$ with $\lambda=0$, 0.05, and 0.1. The magenta area is the central region, the black line is the median, the red lines are outliers, and the green dash line is the reference for separability in theory.}
\end{figure}

We choose five different values of $\lambda=0$, $0.025$, $0.05$, $0.075$, and $0.1$ in the Gneiting model $C^{G'}(\bh,u;\lambda)$ to generate observations $\mathbf{Z}$. The reason for choosing relatively small values for $\lambda$ is that the power of the rank-based test is already strong enough for small values of $\lambda$.
Three examples of the estimated symmetry test functions are shown in Figure~\ref{fig:symmSim}. Since the chosen values of $\lambda$ are very small, the deviation of the symmetry test functions from zero in our most extreme case when $\lambda=0.1$ is still subtle; see Figure~\ref{fig:symm}. 
Although the functional medians of the estimated test functions do not show obvious changes for the different values of $\lambda$, we can still see that the majority of the symmetry test functions move away from zero as $\lambda$ increases. Furthermore, the results of the rank-based test in Table~\ref{tab:symmPower} show that the power is very strong, even when $\lambda$ is as small as $0.01$, and the Type I errors are all close to the nominal levels.

\begin{table}[h!]
	\centering
	\caption{\label{tab:symmPower}Type I error and power for simulated data from the Gneiting model $C^{G'}(\bh,u;\lambda)$ with different values of $\lambda$ in $100$ experiments.}
	\small{
		\begin{tabular}{c|c|c|c|c|c|c|}
			&\begin{tabular}{c}Nominal\\ Level\end{tabular} &Type I error&\multicolumn{4}{|c|}{Power}\\\cline{2-7}
			&$\alpha$&$\lambda=0$&$\lambda=0.025$&$\lambda=0.05$&$\lambda=0.075$&$\lambda=0.1$\\\hline
			Proportion of&$5\%$&0.04&0.07&0.20&0.47&0.70\\\cline{2-7}
			rejections&$1\%$&0.00&0.01&0.07&0.25&0.39\\\hline	
		\end{tabular}
	}
\end{table}

\section{Applications}
\label{sec:application}
\subsection{Wind speed}
\label{subsec:wind}
\begin{figure}[ht!]
	\centering
	\includegraphics[width=0.4\textwidth]{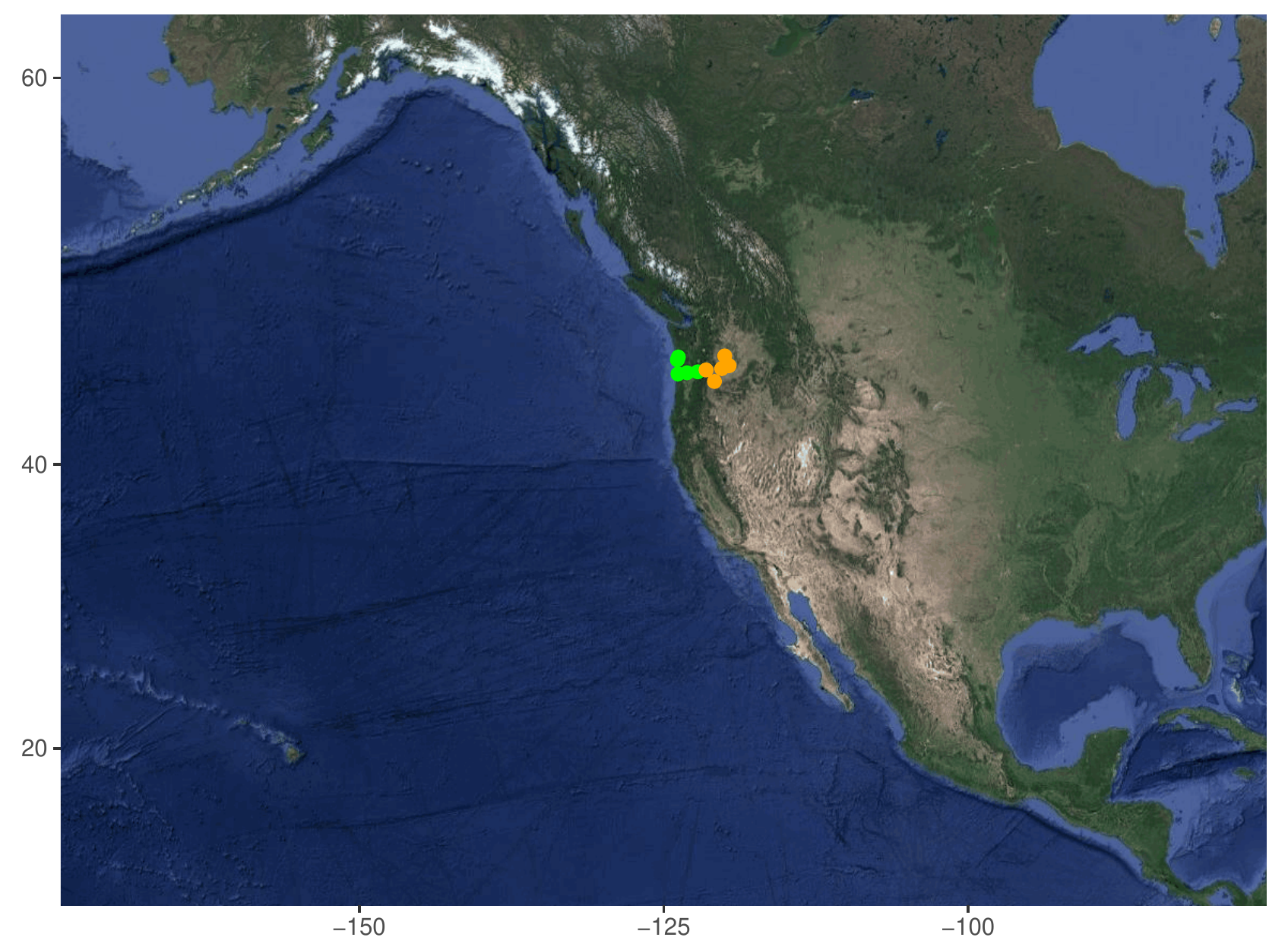}
	\includegraphics[width=0.4\textwidth]{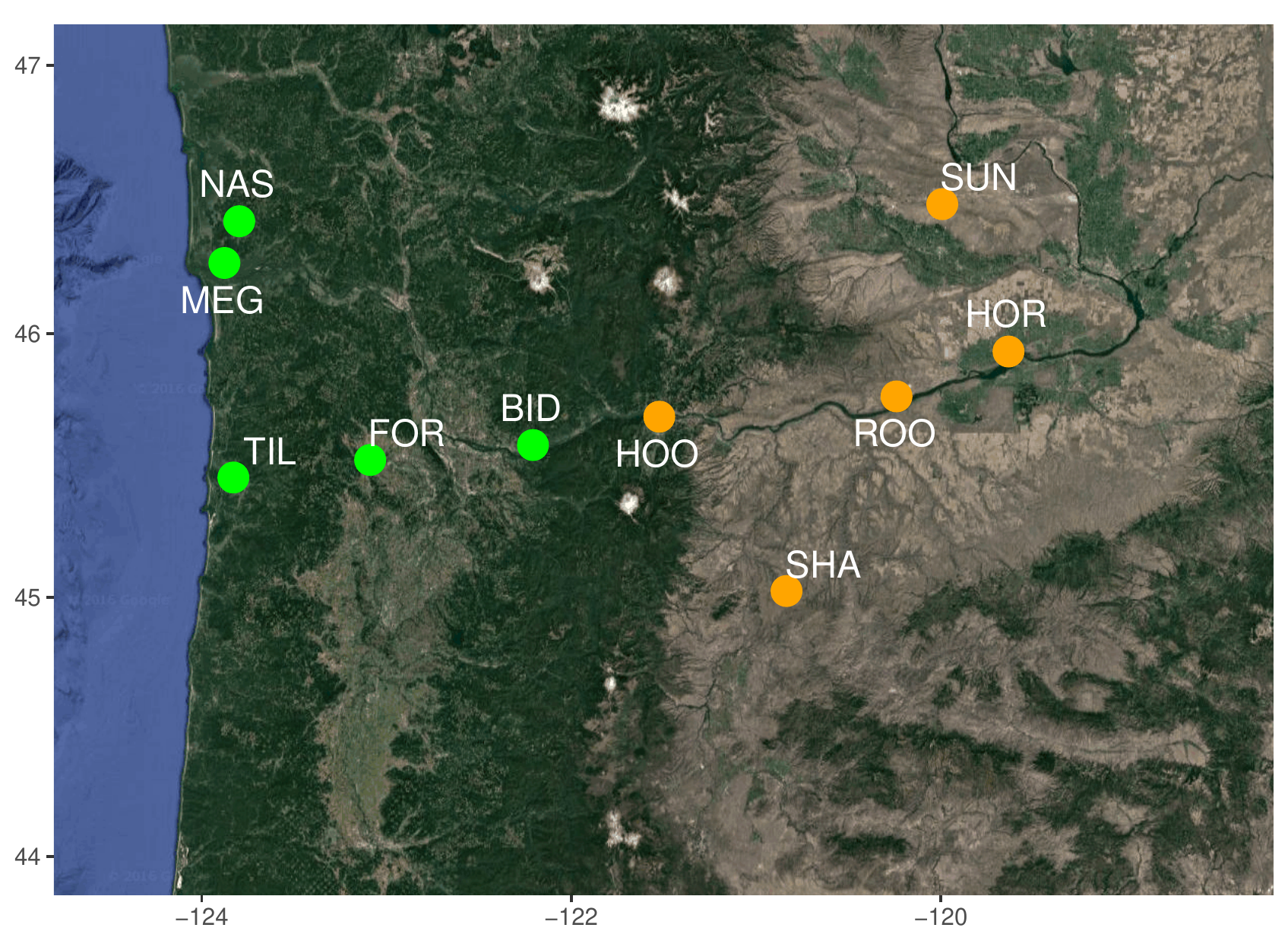}
	\caption{\label{fig:map}Ten locations for observing wind speed in a broad (left) and zoomed (right) view. The green points are grouped in the coastal class and the orange points are grouped in the inland class.}
\end{figure}

Wind speed is an important atmospheric variable in weather forecasting and many other environmental applications. One source of wind speed data is from monitoring stations. In this section, we use our proposed method to visualize and assess properties of covariances in the wind speed from ten monitoring stations in the northwestern U.S.A. The locations are shown in Figure~\ref{fig:map}, where we divide them into two regions. Coastal sites are in green, and inland sites are in orange. We analyze the hourly data observed at each station from 2012 to 2013 and visualize the spatio-temporal covariance structure from two seasons, summer (June to August) and winter (December to February).

\begin{figure}[ht!]
	\centering
	\includegraphics[width=\textwidth]{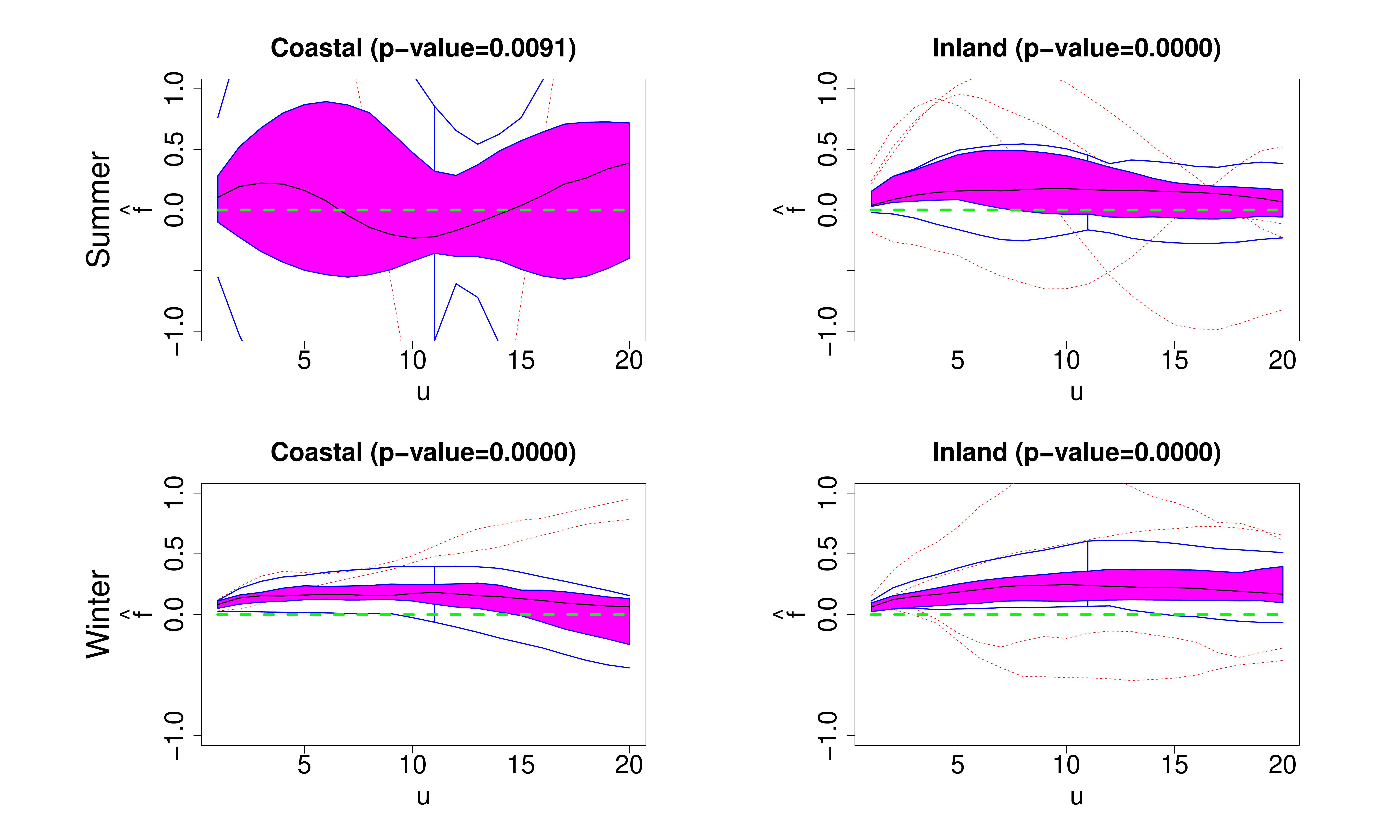}
	\caption{\label{fig:wind}The estimated separability test functions for wind speed in the northwestern U.S.A. during summer and winter, in coastal and inland regions. The p-values of the rank-based test for separability are indicated in the title of each case.}
\end{figure}

\begin{figure}[ht!]
	\centering
	\includegraphics[width=\textwidth]{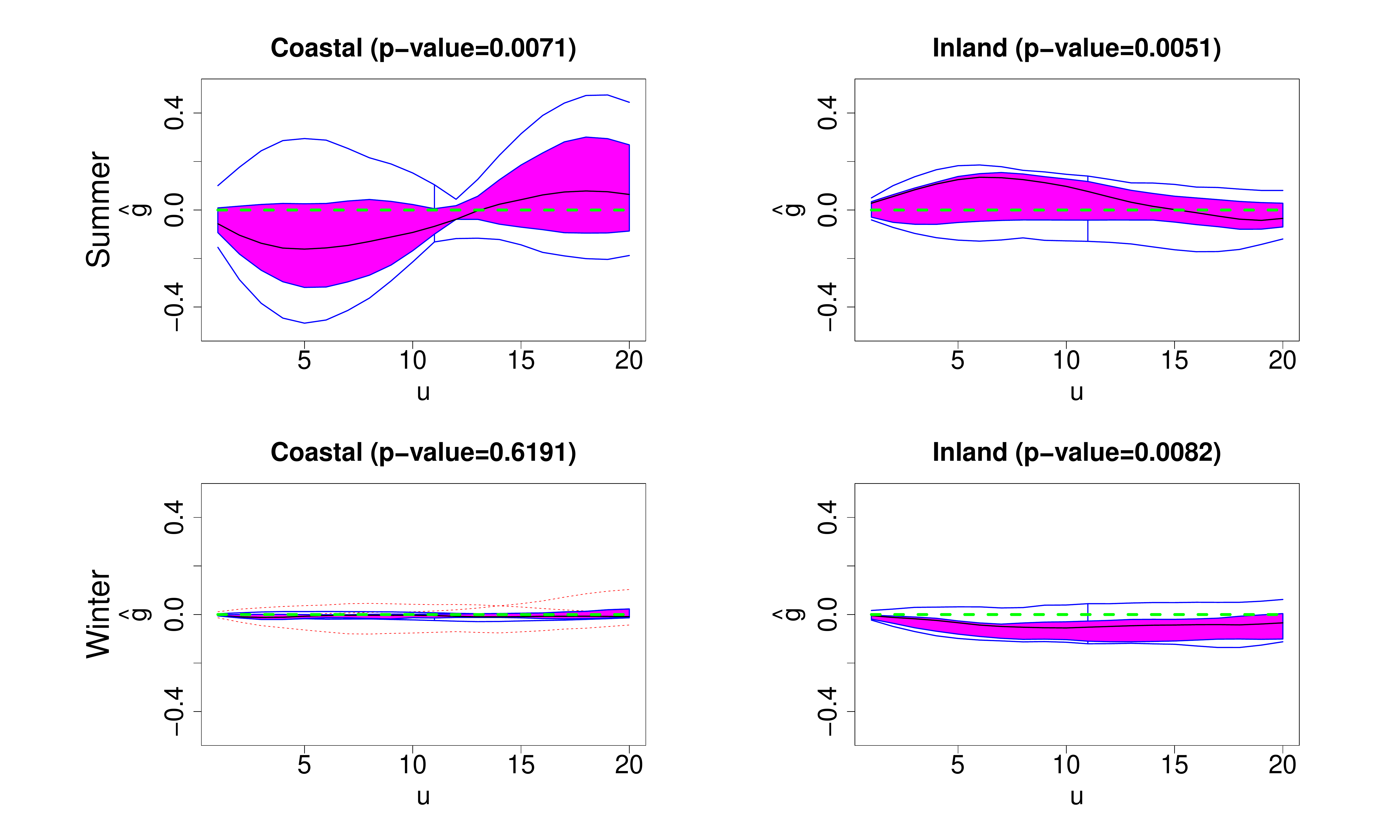}
	\caption{\label{fig:windSymm}The estimated symmetry test functions for wind speed in the northwestern U.S.A. during summer and winter, in coastal and inland regions. The p-values of the rank-based test for symmetry are indicated in the title of each case.}
\end{figure}

The plots of the estimated separability test functions are shown in Figure~\ref{fig:wind}, and the p-value results of the rank-based test are given in the titles. We see the separability test functions are away from zero in all cases, and all the p-values for testing separability are much smaller than $0.05$, indicating rejection of separability at $\alpha=0.05$. When computing the p-value, we choose $10,000$ replicates from the bootstrap to approximate the distribution of the test statistics under the null hypothesis.
Since the covariance functions for all the four cases are significantly non-separable, we next investigate their properties of symmetry.
The symmetry test functions as well as the p-values of the test for different cases are shown in Figure~\ref{fig:windSymm}. The functional boxplot for the coastal region in winter suggests the strongest symmetry, and its corresponding p-value is indeed the largest, while p-values for the other cases are much smaller than $0.05$. 
One possible reason might be that there is no prevailing wind direction during the winter in the coastal region.

\subsection{Climate model outputs}
\begin{figure}
	\centering
	\includegraphics[width=\textwidth]{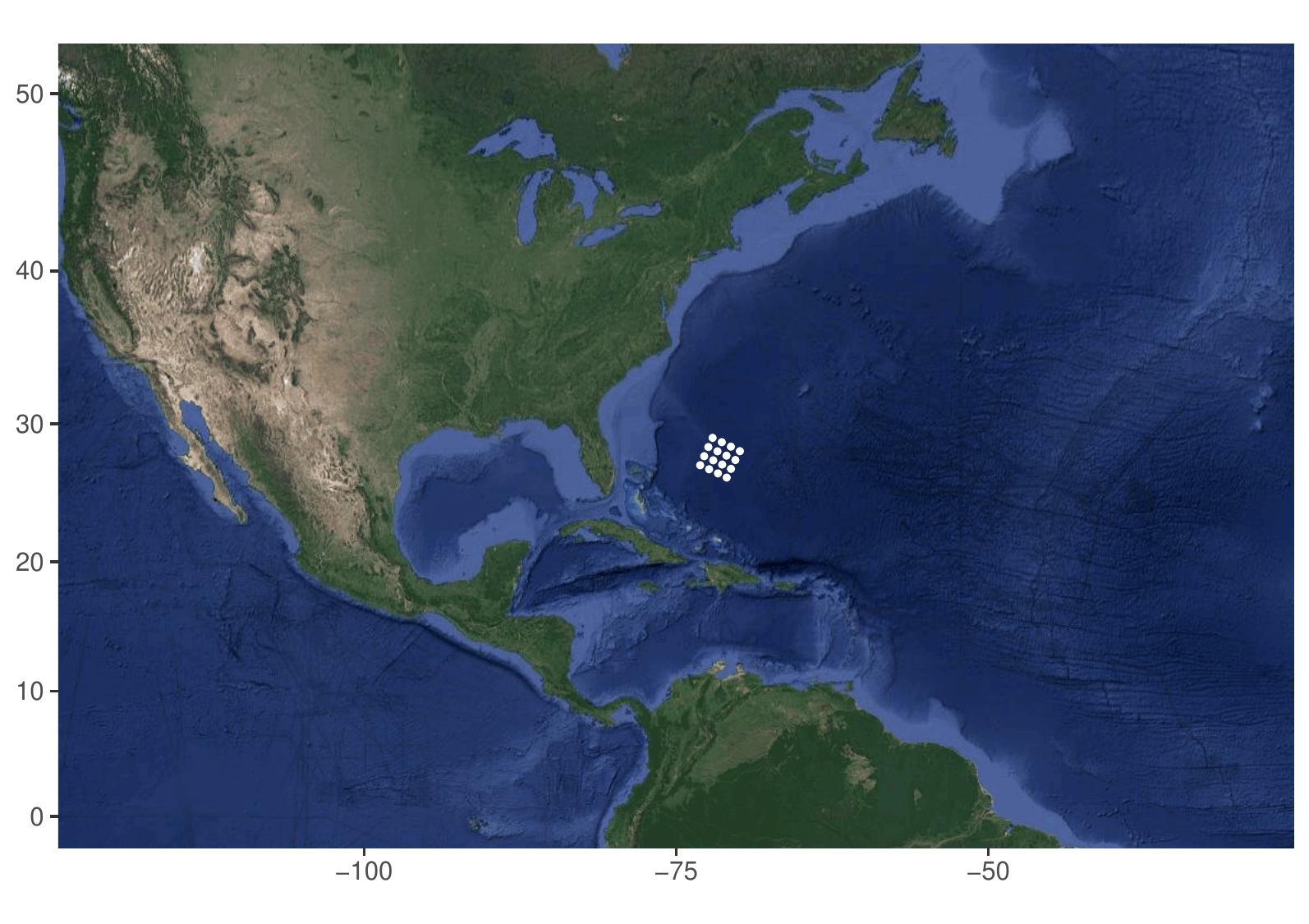}
	\caption{\label{fig:GCMRCMmap}Study area in the North Atlantic Ocean, where the white dots indicate the spatial locations from the model outputs.}
\end{figure}

In climate studies, numerical models are used to simulate many important variables by solving a set of dynamic equations. Two types of models are available: General Circulation Model (GCM) is used to describe the dynamic system of the entire earth, with a global coverage but a comparatively low resolution; however, Regional Climate Model (RCM) with GCM outputs as their boundary conditions is used to simulate locally high-resolution data. Different combinations of GCMs and RCMs may lead to different results. In this section, we apply our visualization and assessment tools to data based on four combinations of two GCMs,  GFDL and HADCM3, and two RCMs,  ECP2 and HRM3, which are provided by the North American Regional Climate Change Assessment Program (NARCCAP)~\citep{Mearns2012}. More details about these numerical models can be found at \url{http://www.narccap.ucar.edu/}. We focus on a small region which consists of $16$ spatial locations in the North Atlantic Ocean (as shown in Figure~\ref{fig:GCMRCMmap}) and study the daily surface temperature and wind speed from 1976 to 1979.

\begin{figure}
	\centering
	\includegraphics[width=\textwidth]{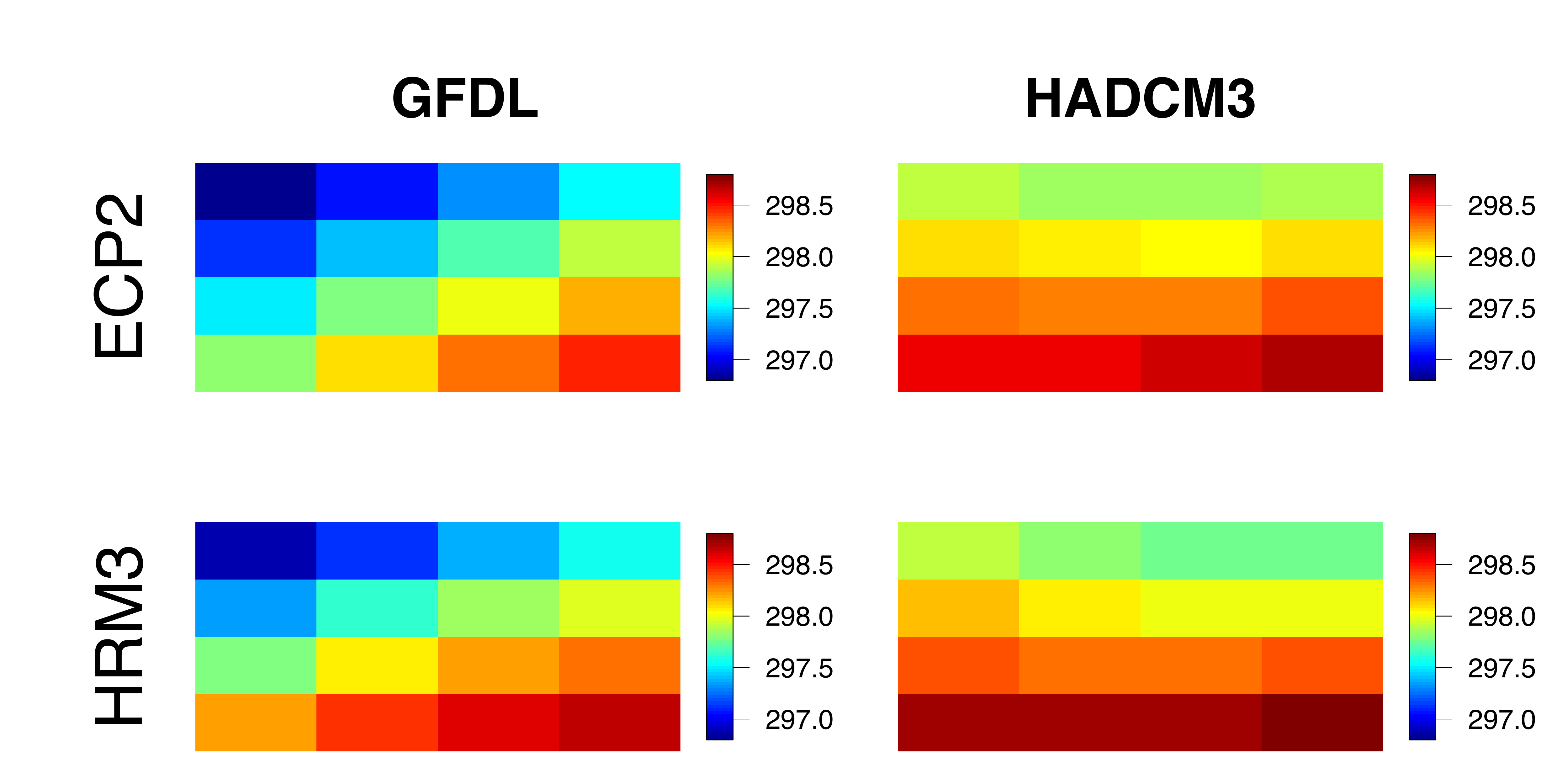}
	\caption{\label{fig:GCMRCMmean}Mean of the sea surface temperature (unit:~K) during 1976--1979 at each of the $16$ locations.}
\end{figure}

\begin{figure}
	\centering
	\includegraphics[width=\textwidth]{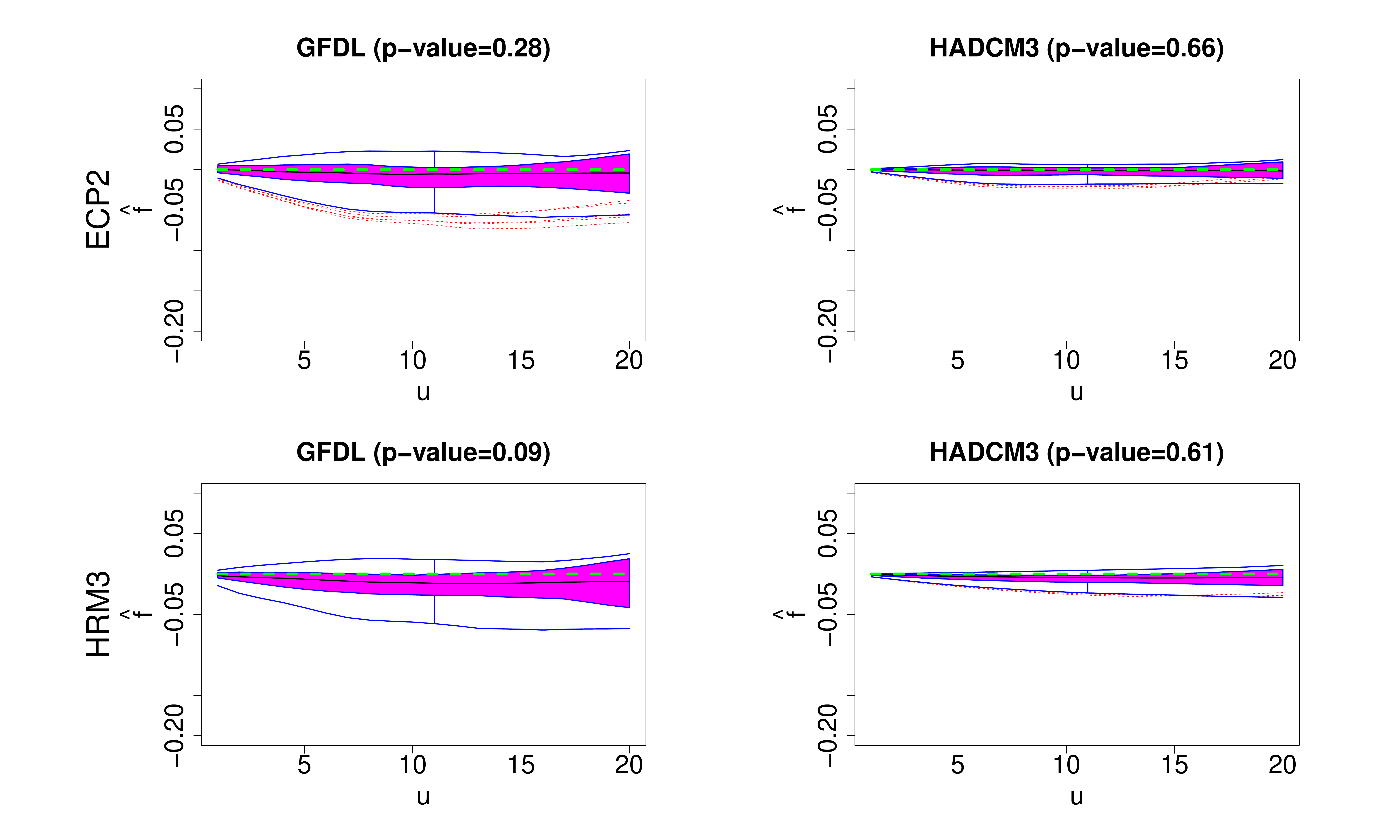}
	\caption{\label{fig:GCMRCMsep}The estimated separability test functions for daily surface temperature with different combinations of GCMs and RCMs. The p-values of the rank-based test for separability are indicated in the title of each case.}
\end{figure}

We first look at the mean of the surface temperature across the four years at each location, which is shown in Figure~\ref{fig:GCMRCMmean}. We see different GCMs as boundary conditions give different patterns of surface temperatures, while different RCMs show similar results. 
We then apply our visualization and assessment tools to the covariance structures of these four GCM and RCM combinations.
To eliminate seasonality, we remove the monthly mean from the daily temperatures. The estimated separability test functions and the p-values of the rank-based test for separability are shown in Figure~\ref{fig:GCMRCMsep}.
With all the p-values greater than $0.05$, there is no significant evidence that the covariances of these four cases are non-separable at the $5\%$ significance level. 
However, the output from GFDL and HRM3 shows stronger non-separability and the separability is rejected at the $10\%$ significance level with a p-value $0.09$.

\begin{figure}
	\centering
	\includegraphics[width=\textwidth]{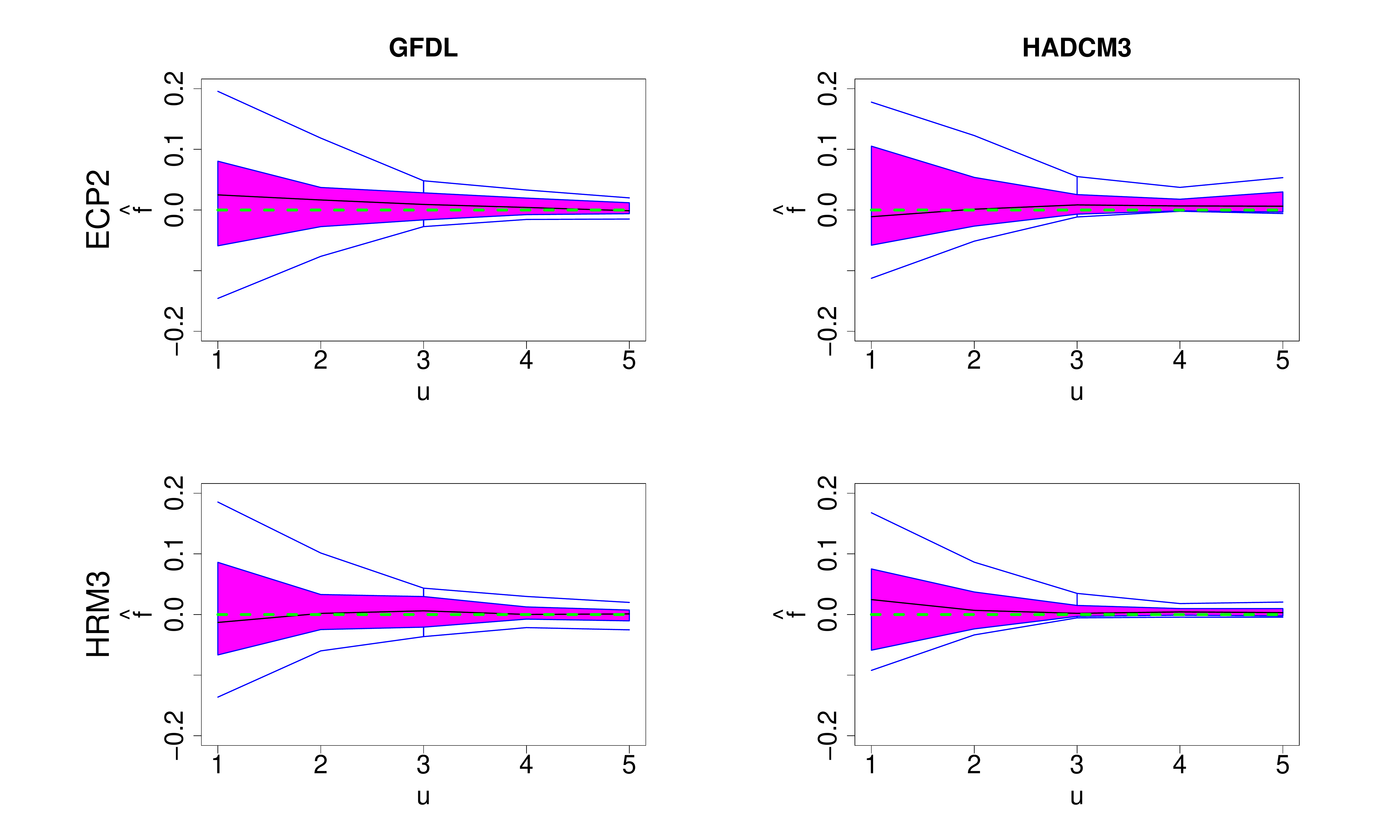}
	\caption{\label{fig:GCMRCMwindSep}The estimated separability test functions for daily wind speed with different combinations of GCMs and RCMs.}
\end{figure}

\begin{figure}
	\centering
	\includegraphics[width=\textwidth]{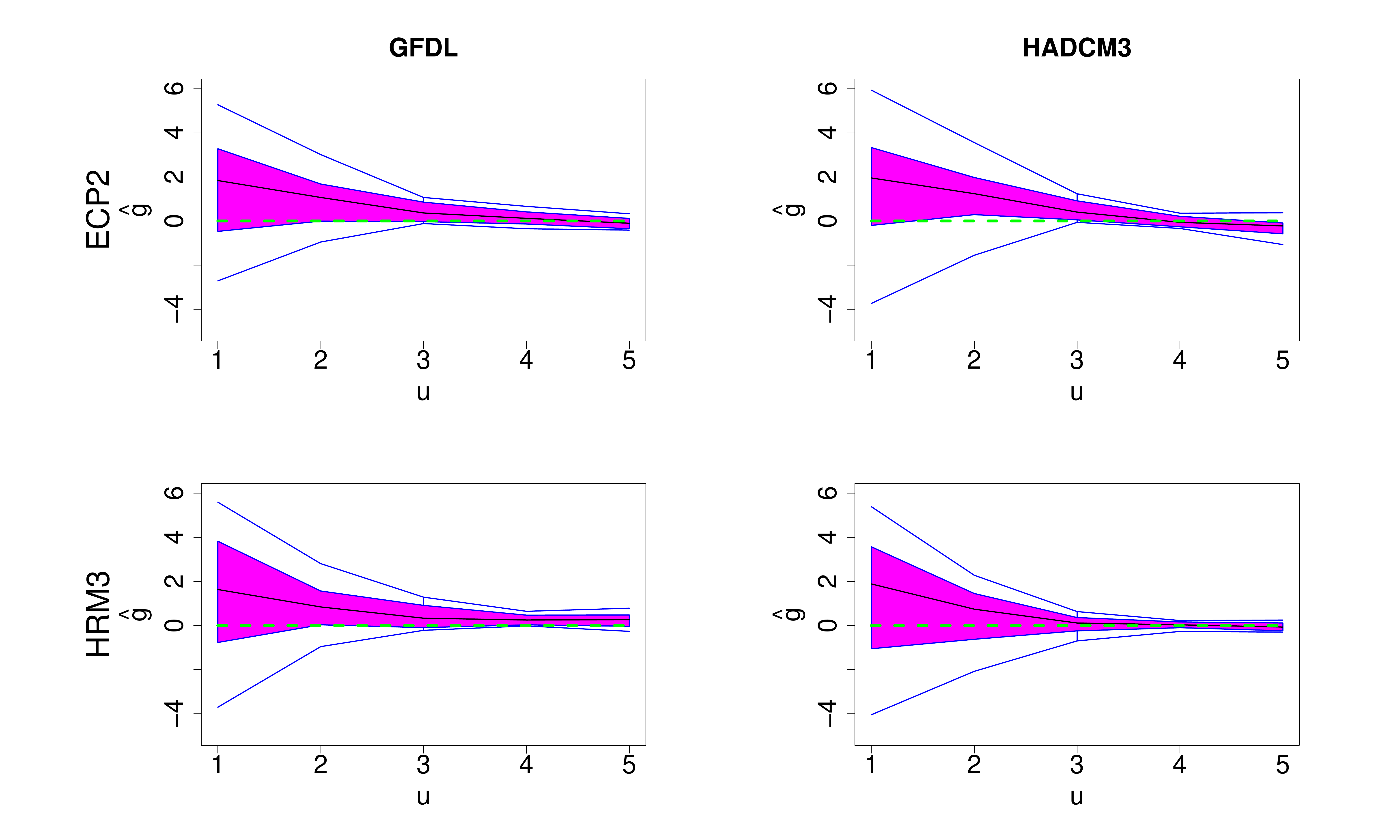}
	\caption{\label{fig:GCMRCMwindSymm}The estimated symmetry test functions for daily wind speed with different combinations of GCMs and RCMs.}
\end{figure}

Next, we look at the daily wind speed in the same area.
The mean wind speeds at these locations are all close to zero and do not show much difference. Again, we remove the monthly mean from the daily wind speed and then apply our visualization and assessment methods. 
The estimated separability test functions are shown in Figure~\ref{fig:GCMRCMwindSep}, where the medians are away from zero.
These test functions show large variability in the functional boxplot, and all the p-values from the rank-based tests are as small as zero, indicating strong separability. 
To further investigate the symmetry, we plot the estimated symmetry test functions; the illustration is given in Figure~\ref{fig:GCMRCMwindSymm}. All the symmetry test functions are also very far from zero, and we find that the p-values of the test for symmetry are zero. Therefore, we conclude that the covariance of wind speed is neither separable nor symmetric.

In this application, the covariance of the daily surface temperatures tends to be separable and thus also symmetric; for daily wind speed, the covariance is non-separable and asymmetric.
Our analysis suggests that, for the region we have considered, the daily surface temperatures produced by the climate models do not show significant space-time interaction, whereas the daily wind speeds clearly show asymmetric space-time interaction.

\section{Discussion}
We presented a functional data analysis approach to visualizing and assessing spatio-temporal covariance properties.
The proposed method is suitable for visualizing the separability and the symmetry of a stationary spatio-temporal process. The p-value calculated in the rank-based test can be used to measure the degree of separability or symmetry. We illustrated our methods using various classes of spatio-temporal covariance models, and our simulations demonstrated the good performance of our proposed methods. 
In the practical applications, we applied our method to temperature and wind speed data from either monitoring sites or climate model outputs, and illustrated how to interpret the separability and symmetry in these different scenarios. 

However, we have only considered spatio-temporal datasets collected from a small number of spatial locations. 
Because we need to estimate the test functions for each pair of locations, computation will become an issue as the number of locations grows.
One possible solution is to divide the region of interest into smaller subregions, and estimate all the test functions separately for each of these subregions. Then, the obtained test functions can be combined for the visualization and the rank-based test. This would significantly reduce the number of test functions, and each small subregion is more likely to be stationary. However, future research is required to assess the performance of this approach.

\section*{\centering Acknowledgement}
We wish to thank the North American Regional Climate Change Assessment Program (NARCCAP) for providing the data used in this paper. NARCCAP is funded by the National Science Foundation (NSF), the U.S. Department of Energy (DoE), the National Oceanic and Atmospheric Administration (NOAA), and the U.S. Environmental Protection Agency Office of Research and Development (EPA).
	
\newpage
\bibliographystyle{chicago}
\bibliography{reference}

\begin{thebibliography}{}

\bibitem[\protect\citeauthoryear{Brown, Diggle, Lord, and Young}{Brown
  et~al.}{2001}]{Brown2001}
Brown, P.~E., P.~J. Diggle, M.~E. Lord, and P.~C. Young (2001).
\newblock {Space-time calibration of radar rainfall data}.
\newblock {\em Journal of the Royal Statistical Society Series C Applied
  Statistics\/}~{\em 50\/}(2), 221--241.

\bibitem[\protect\citeauthoryear{Cesare, Myers, and Posa}{Cesare
  et~al.}{2001}]{Cesare2001}
Cesare, L.~D., D.~E. Myers, and D.~Posa (2001).
\newblock {Estimating and modeling space-time correlation structures}.
\newblock {\em Statistics and Probability Letters\/}~{\em 51\/}(1), 9--14.

\bibitem[\protect\citeauthoryear{Cressie and Huang}{Cressie and
  Huang}{1999}]{Cressie1999}
Cressie, N. and H.~Huang (1999).
\newblock {Classes of nonseparable, spatio-temporal stationary covariance
  functions}.
\newblock {\em Journal of the American Statistical Association\/}~{\em 94},
  1330--1340.

\bibitem[\protect\citeauthoryear{Fraiman and Muniz}{Fraiman and
  Muniz}{2001}]{Fraiman2001}
Fraiman, R. and G.~Muniz (2001).
\newblock {Trimmed means for functional data}.
\newblock {\em Sociedad de Estadistica e Investigacion Operativa\/}~{\em
  10\/}(2), 419--440.

\bibitem[\protect\citeauthoryear{Fuentes}{Fuentes}{2006}]{Fuentes2006}
Fuentes, M. (2006).
\newblock {Testing for separability of spatial-temporal covariance functions}.
\newblock {\em Journal of Statistical Planning and Inference\/}~{\em 136\/}(2),
  447--466.

\bibitem[\protect\citeauthoryear{Gneiting}{Gneiting}{2002}]{Gneiting2002}
Gneiting, T. (2002).
\newblock {Nonseparable, stationary covariance functions for space–time
  data}.
\newblock {\em Journal of the American Statistical Association\/}~{\em
  97\/}(458), 590--600.

\bibitem[\protect\citeauthoryear{Gneiting, Genton, and Guttorp}{Gneiting
  et~al.}{2006}]{Gneiting2007}
Gneiting, T., M.~G. Genton, and P.~Guttorp (2006).
\newblock {Geostatistical space-time models, stationarity, separability, and
  full symmetry}.
\newblock {\em Monographs On Statistics and Applied Probability\/}, 151--175.

\bibitem[\protect\citeauthoryear{Jones and Zhang}{Jones and
  Zhang}{1997}]{Jones1997}
Jones, R.~H. and Y.~Zhang (1997).
\newblock {Models for continuous stationary space-time processes}.
\newblock In T.~G. Gregoire, D.~R. Brillinger, P.~J. Diggle, E.~Russek-Cohen,
  W.~G. Warren, and R.~D. Wolfinger (Eds.), {\em Modelling Longitudinal and
  Spatially Correlated Data}, pp.\  289--298. New York, NY: Springer New York.

\bibitem[\protect\citeauthoryear{Kyriakidis and Journel}{Kyriakidis and
  Journel}{1999}]{Kyriakidis1999}
Kyriakidis, P.~C. and A.~G. Journel (1999).
\newblock {Geostatistical space–time models: a review}.
\newblock {\em Mathematical Geology\/}~{\em 31\/}(6), 651--684.

\bibitem[\protect\citeauthoryear{Li, Genton, and Sherman}{Li
  et~al.}{2007}]{Li2007}
Li, B., M.~G. Genton, and M.~Sherman (2007).
\newblock {A nonparametric assessment of properties of space–time covariance
  functions}.
\newblock {\em Journal of the American Statistical Association\/}~{\em
  102\/}(478), 736--744.

\bibitem[\protect\citeauthoryear{L{\'{o}}pez-Pintado and
  Romo}{L{\'{o}}pez-Pintado and Romo}{2009}]{Lopez-Pintado2009}
L{\'{o}}pez-Pintado, S. and J.~Romo (2009).
\newblock {On the concept of depth for functional data}.
\newblock {\em Journal of the American Statistical Association\/}~{\em
  104\/}(486), 718--734.

\bibitem[\protect\citeauthoryear{L{\'{o}}pez-Pintado and
  Romo}{L{\'{o}}pez-Pintado and Romo}{2011}]{Lopez-Pintado2011}
L{\'{o}}pez-Pintado, S. and J.~Romo (2011).
\newblock {A half-region depth for functional data}.
\newblock {\em Computational Statistics and Data Analysis\/}~{\em 55\/}(4),
  1679--1695.

\bibitem[\protect\citeauthoryear{Mearns, Arritt, Biner, Bukovsky, McGinnis,
  Sain, Caya, Correia, Flory, Gutowski, Takle, Jones, Leung, Moufouma-Okia,
  McDaniel, Nunes, Qian, Roads, Sloan, and Snyder}{Mearns
  et~al.}{2012}]{Mearns2012}
Mearns, L.~O., R.~Arritt, S.~Biner, M.~S. Bukovsky, S.~McGinnis, S.~Sain,
  D.~Caya, J.~Correia, D.~Flory, W.~Gutowski, E.~S. Takle, R.~Jones, R.~Leung,
  W.~Moufouma-Okia, L.~McDaniel, A.~M.~B. Nunes, Y.~Qian, J.~Roads, L.~Sloan,
  and M.~Snyder (2012).
\newblock {The North American regional climate change assessment program:
  overview of phase I results}.
\newblock {\em Bulletin of the American Meteorological Society\/}~{\em
  93\/}(9), 1337--1362.

\bibitem[\protect\citeauthoryear{Mitchell, Genton, and Gumpertz}{Mitchell
  et~al.}{2005}]{Mitchell2005}
Mitchell, M.~W., M.~G. Genton, and M.~L. Gumpertz (2005).
\newblock {Testing for separability of space-time covariances}.
\newblock {\em Environmetrics\/}~{\em 16\/}(8), 819--831.

\bibitem[\protect\citeauthoryear{Mitchell, Genton, and Gumpertz}{Mitchell
  et~al.}{2006}]{Mitchell2006}
Mitchell, M.~W., M.~G. Genton, and M.~L. Gumpertz (2006).
\newblock {A likelihood ratio test for separability of covariances}.
\newblock {\em Journal of Multivariate Analysis\/}~{\em 97\/}(5), 1025--1043.

\bibitem[\protect\citeauthoryear{Narisetty and Nair}{Narisetty and
  Nair}{2016}]{Narisetty2015}
Narisetty, N.~N. and V.~N. Nair (2016).
\newblock {Extremal depth for functional data and applications}.
\newblock {\em Journal of the American Statistical Association\/}~{\em
  111\/}(516), 1705--1714.

\bibitem[\protect\citeauthoryear{Rodrigues and Diggle}{Rodrigues and
  Diggle}{2010}]{Rodrigues2010}
Rodrigues, A. and P.~J. Diggle (2010).
\newblock {A class of convolution‐based models for spatio‐temporal
  processes with non‐separable covariance structure}.
\newblock {\em Scandinavian Journal of Statistics\/}~{\em 37\/}(4), 553--567.

\bibitem[\protect\citeauthoryear{Scaccia and Martin}{Scaccia and
  Martin}{2005}]{Scaccia2005}
Scaccia, L. and R.~J. Martin (2005).
\newblock {Testing axial symmetry and separability of lattice processes}.
\newblock {\em Journal of Statistical Planning and Inference\/}~{\em 131\/}(1),
  19--39.

\bibitem[\protect\citeauthoryear{Shitan and Brockwell}{Shitan and
  Brockwell}{1995}]{Shitan1995}
Shitan, M. and P.~J. Brockwell (1995).
\newblock {An asymptotic test for separability of a spatial autoregressive
  model}.
\newblock {\em Communications in Statistics - Theory and Methods\/}~{\em
  24\/}(8), 2027--2040.

\bibitem[\protect\citeauthoryear{Stein}{Stein}{2005}]{Stein2005}
Stein, M.~L. (2005).
\newblock {Space–time covariance functions}.
\newblock {\em Journal of the American Statistical Association\/}~{\em
  100\/}(469), 310--321.

\bibitem[\protect\citeauthoryear{Sun and Genton}{Sun and
  Genton}{2011}]{Sun2011}
Sun, Y. and M.~G. Genton (2011).
\newblock {Functional boxplots}.
\newblock {\em Journal of Computational and Graphical Statistics\/}~{\em
  20\/}(2), 316--334.

\bibitem[\protect\citeauthoryear{Sun and Genton}{Sun and
  Genton}{2012a}]{Sun2012a}
Sun, Y. and M.~G. Genton (2012a).
\newblock {Adjusted functional boxplots for spatio-temporal data visualization
  and outlier detection}.
\newblock {\em Environmetrics\/}~{\em 23\/}(1), 54--64.

\bibitem[\protect\citeauthoryear{Sun and Genton}{Sun and
  Genton}{2012b}]{Sun2012b}
Sun, Y. and M.~G. Genton (2012b).
\newblock {Functional median polish}.
\newblock {\em Journal of Agricultural, Biological, and Environmental
  Statistics\/}~{\em 17\/}(3), 354--376.

\end{thebibliography}

\end{document}